\documentclass[12pt, letterpaper]{article}

\usepackage{graphicx}
\usepackage{amssymb}
\usepackage{amsmath}
\usepackage[figuresright]{rotating}
\usepackage{bm}
\usepackage[utf8]{inputenc}
\usepackage{mathtools}
\usepackage{cite}
\usepackage{hyperref}

\title{\fontsize{14}{15}\bfseries Deep Reinforcement Learning for Cybersecurity Threat Detection and Protection: A Review}

\author{
Mohit Sewak\\
\texttt{Microsoft R\&D, India}\\
\texttt{mohit.sewak@microsoft.com}
\and
Sanjay K. Sahay, Hemant Rathore\\
\texttt{BITS Pilani, Goa, India}\\
\texttt{\{ssahay, hemantr\}@goa.bits-pilani.ac.in}
}

\date{} 

\begin{document}
\maketitle

\begin{abstract}
The cybersecurity threat landscape has lately become overly complex. Threat actors leverage weaknesses in the network and endpoint security in a very coordinated manner to perpetuate sophisticated attacks that could bring down the entire network and many critical hosts in the network. Increasingly advanced deep and machine learning-based solutions have been used in threat detection and protection. The application of these techniques has been reviewed well in the scientific literature. Deep Reinforcement Learning has shown great promise in developing AI-based solutions for areas that had earlier required advanced human cognizance. Different techniques and algorithms under deep reinforcement learning have shown great promise in applications ranging from games to industrial processes, where it is claimed to augment systems with general AI capabilities. These algorithms have recently also been used in cybersecurity, especially in threat detection and endpoint protection, where these are showing state-of-the-art results. Unlike supervised machine and deep learning, deep reinforcement learning is used in more diverse ways and is empowering many innovative applications in the threat defense landscape. However, there does not exist any comprehensive review of these unique applications and accomplishments. Therefore, in this paper, we intend to fill this gap and provide a comprehensive review of the different applications of deep reinforcement learning in cybersecurity threat detection and protection.

\end{abstract}

\section{Introduction}\label{sec:intro}
With the exponential rise in data, the need for its protection from theft and damage has become particularly important. The modern-day attacks are much more sophisticated, and when paired with the rise in cloud services, smartphones, and the Internet of Things (IoT) devices, we now have a complex defense scenario amidst a myriad of new cybersecurity threats that did not exist a few decades ago. Coordinated attacks that initially intrude via the network layer, and then infect multiple hosts in quick successions are not uncommon even for non-state/military networks and hosts now. Therefore, a robust Managed (Threat) Detection and Response (MDR) system is necessary to provide integrated security for the network and endpoint from threats and malicious activities.

Deep Reinforcement Learning (DRL) is gaining popularity in various fields ranging from games to industrial processes and cyber-physical systems. It has recently started gaining traction in various aspects of cybersecurity as well. Though there exist surveys on use of DRL in security \cite{drl-security-nguyen}, but their focus is not MDR. Therefore, in this paper, we have presented a review of the various applications of DRL-based techniques and how they have improved various aspects of MDR. As shown in figure \ref{fig:mdr-taxonomy}, the MDR integrates two important threat detection and prevention systems as follows:

\begin{itemize}
    \item Intrusion (Detection \&) Prevention System (IDPS/IPS): an IDPS/IPS is an intrusion detection and prevention system. The IDPS/IPS works in conjunction with one or more Intrusion Detection Systems (IDS).
    \item Endpoint Detection \& Response system (EDR): The role of an EDR system is to secure the endpoints. Modern EDR also integrates the Host IDS (HIDS), besides the Endpoint Protection Platform (EPP), and Advanced Threat Protection (ATP).
\end{itemize}

The market for EDR solutions is growing at a rapid pace, from \$238 Mn. in 2015 to \$1.54 Bn. in 2020 while that of the global Intrusion Detection And Prevention Systems (IDPS) is projected to grow by 5.4\% (from USD 4.8 Bn. in 2020 to USD 6.2 Bn. by 2025) \cite{statistica-epp-edr}. The major factors responsible for these changes include the increasing number of attacks, the rising privacy and security awareness, etc. Hence the need to use the best technologies and frameworks to keep up with the rapidly growing market is necessary. DRL offers advanced solutions for many of the needs of both IDS and EPP, and with the increasing funding trends in advanced Artificial Intelligence (AI) based solutions in this area, we believe that a review of the different applications of DRL in various aspects of threat detection and protection is very much needed at this stage.

The rest of the paper is organised as follows. In section \ref{sec:about-drl} we provide a brief introduction on the diverse types of DRL techniques. Next, we provide a similar introduction of the MDR, IDS and EDR systems, along with their respective taxonomies in section \ref{sec:about-threat-detection-systems}. Next, we describe and analyse the different arts that use DRL in the IDS space in section \ref{sec:drl-nids}, and in the endpoint space in section \ref{sec:drl-endpoint}. Finally we conclude the paper in section \ref{sec:conclusion} .

\section{About Deep Reinforcement Learning}\label{sec:about-drl}
\begin{figure*}[tb!]
    \centering
    \includegraphics[width=\textwidth]{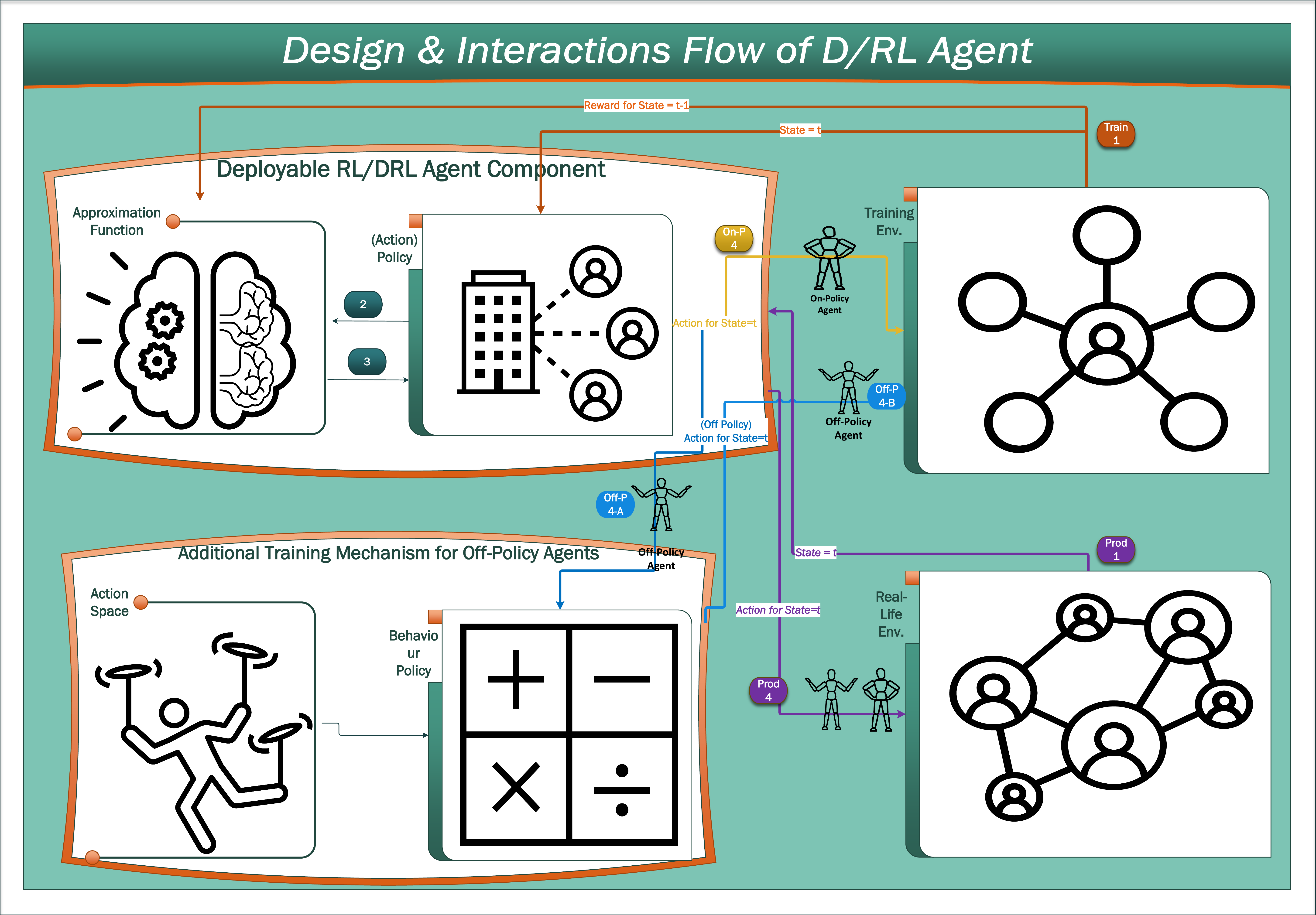}
    \caption{Interaction of different types of D/RL Agents}
    \label{fig:drl-interaction-types}
\end{figure*}

Reinforcement learning (RL) \cite{drl-1-intro-rl} is the field of machine learning (ML) that deals with sequential decision-making involving an agent which learns the desired action policy (behavior) incrementally while interacting with the environment. RL explores the different states of an environment using an explore-exploit mechanism/policy and hence the RL agent does not require complete knowledge or control of the environment. The exploration policy of an off-policy agent could be different from its action policy. Whereas, for an on policy-agent, a stochastic action policy also aids the exploration of the environment. This is as shown in figure \ref{fig:drl-interaction-types}.

\begin{figure*}[htb!]
    \centering
    \includegraphics[width=\textwidth]{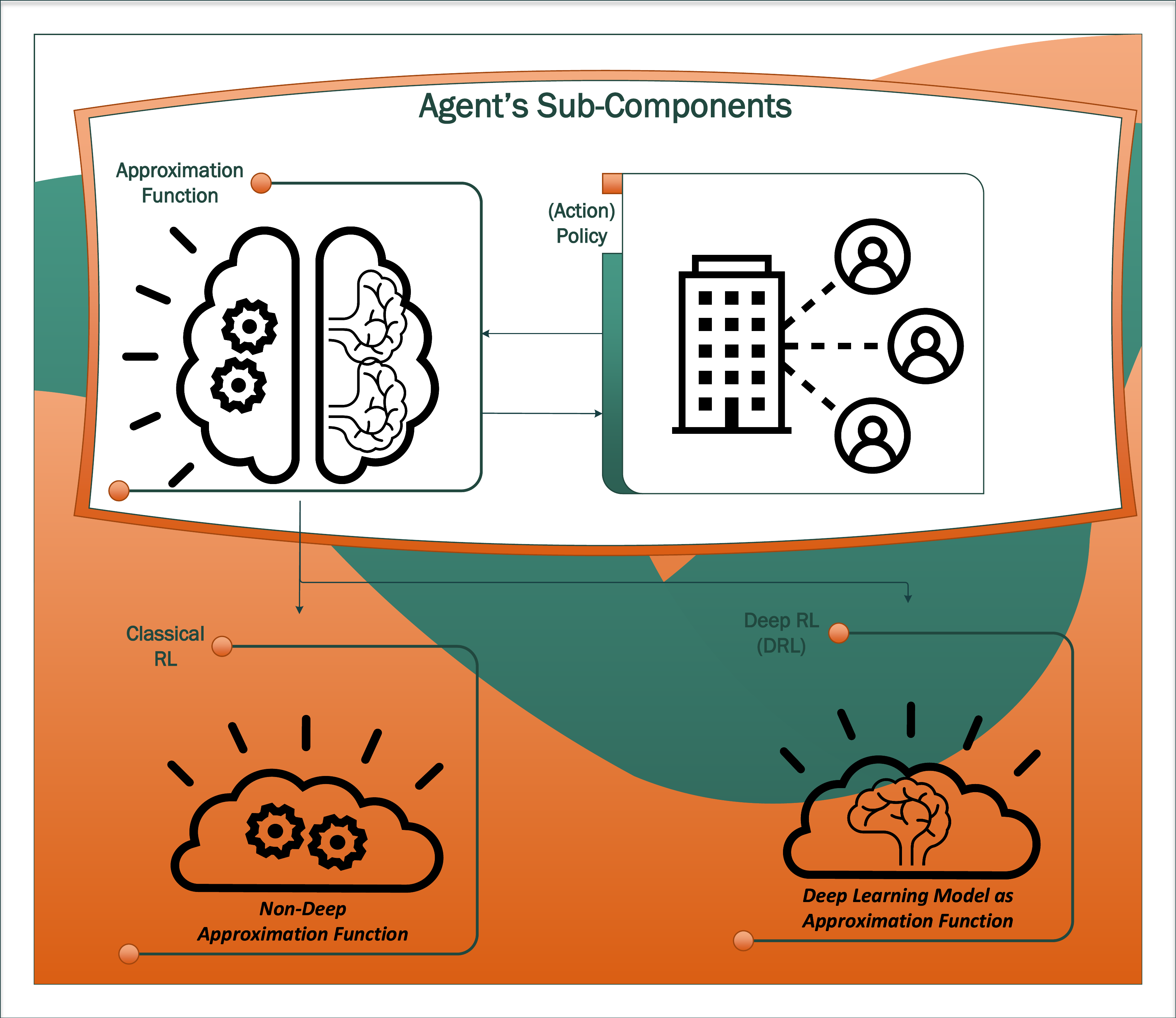}
    \caption{Classical RL vs. Deep RL (DRL)}
    \label{fig:rl-vs-drl}
\end{figure*}

RL is different from any supervised ML or Deep Learning (DL) algorithm. Where supervised ML/DL aims to either maximize a likelihood function or minimize a loss function to estimate/predict an optimal value against a specific record, RL tries to maximize total (absolute or discounted) reward over a trajectory of observations. The observations in an LR context comprises of a set of states, each of these states individually may closely resemble the input of a supervised RL/DL problem. Therefore, an RL/DRL problem may house within itself a supervised learning ML/DL algorithm to provide an estimate against an individual state, which will be helpful for the agent to determine the policy to maximize the overall reward. This is where the distinction between RL (also known as Classical RL) and Deep Reinforcement Learning (DRL) \cite{drl-an-overview-li,drl-survey-arulkumaran} comes in. In RL, the approximation function is either a simple classical ML algorithm or even a non-ML model, like a tabular-memory structure as in the case of Q-Learning; whereas in DRL \cite{Sewak-DRL}, such approximation functions are invariably a DL algorithm. This distinction is illustrated in figure \ref{fig:rl-vs-drl}.

DL-based approximators as used in DRL are well suited for handling and extracting insights from high-dimensional sensory inputs. These algorithms can process states consisting of inputs ranging from raw image frames to complex raw sensor data. Where classical RL needs human expertise to extract domain-relevant insightful features from the input states, the DL model in the DRL algorithm can automatically extract complex non-linear features from raw-input features, thus making them ideal for learning complex and dynamic real-world processes.

\begin{figure*}[tb!]
    \centering
    \includegraphics[width=\textwidth]{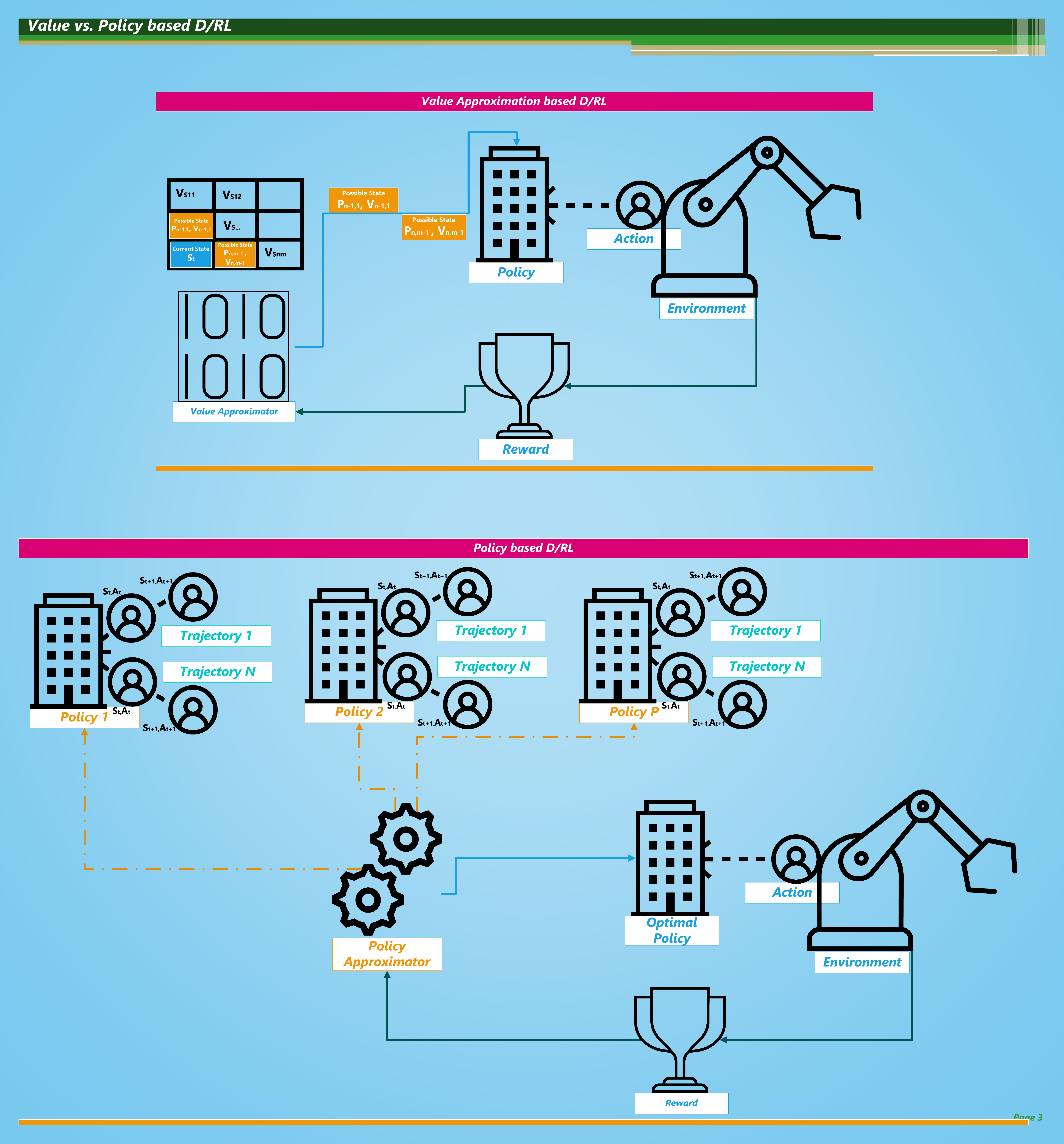}
    \caption{Difference between Value-based and Policy-based D/RL Approaches}
    \label{fig:value-vs-policy-drl}
\end{figure*}

A DRL-based algorithm could be further sub-divided into value approximation-based algorithm \cite{sewak-ValueDRL} or policy-based \cite{drl-10-policy-based-approaches} approaches and algorithm. This distinction is based upon the underlying utility that the approximation function is targeted to estimate. This is illustrated in figure \ref{fig:value-vs-policy-drl}. If the approximation algorithm aims to estimate the value of being in a future state, which further influences the policy of the agent, then such approach and hence the algorithm falls under the value (approximation based) D/RL approaches. The common value based DRL algorithms are as follows \cite{drl-8-dqn-ddqn}: 
\begin{itemize}
    \item Deep Q-networks \cite{DQN_Nature}
    \item Double DQN \cite{Double_DQN}
    \item Dueling DQN (or Dueling network architecture for Q-Learning) \cite{Dueling_DQN}
\end{itemize}

Further there are some not so popular approaches also, like the Distributional DQN, and the AE-DQN \cite{AE-DQN}.

The policy approximation-based methods on the other hand optimize a performance utility function as the objective (typically the expected cumulative reward) by finding a good policy. As compared to estimating the value function that in turn is used to greedily reach an optimal policy as in value-based approaches, the policy-based approach offers a direct solution to optimizing the policy that leads to taking action trajectories that maximize the reward. But since the policy-based approach deals with distributions of complex trajectories instead of simple scalar values, finding a gradient of such distribution and thereby optimizing it becomes challenging. Therefore, although policy-based approaches are more powerful for real-life applications, they are also mathematically and computationally more demanding. The common policy-based DRL algorithms are as follows \cite{drl-10-policy-based-approaches}:

\begin{itemize}
    \item Deep Deterministic Policy Gradient (DDPG) \cite{drl-13-deep-deterministic-policy-gradient}
    \item Trust Region Policy Optimization (TRPO) \cite{TRPO}
    \item Proximal Policy Optimization (PPO) \cite{PPO}
    \item Generalized Advantage Estimation (GAE) \cite{GAE}
\end{itemize}

Most of the DRL approaches covered here are model-free approaches. Classical RL has a vast repository of model-based approaches as well that relies on having complete knowledge of the model of the environment (dynamics and reward function) in conjunction with a planning algorithm. Since the assumption of complete understanding of the environment cannot be fulfilled for large complex environments, popular ML/DL-based RL use approximation functions to estimate the utility of a proxy function under the assumed model of the environment during its exploration.

\section{About Threat Detection and Protection Systems}\label{sec:about-threat-detection-systems}
\begin{figure}[htb!]
    \centering
    \includegraphics[width=\textwidth]{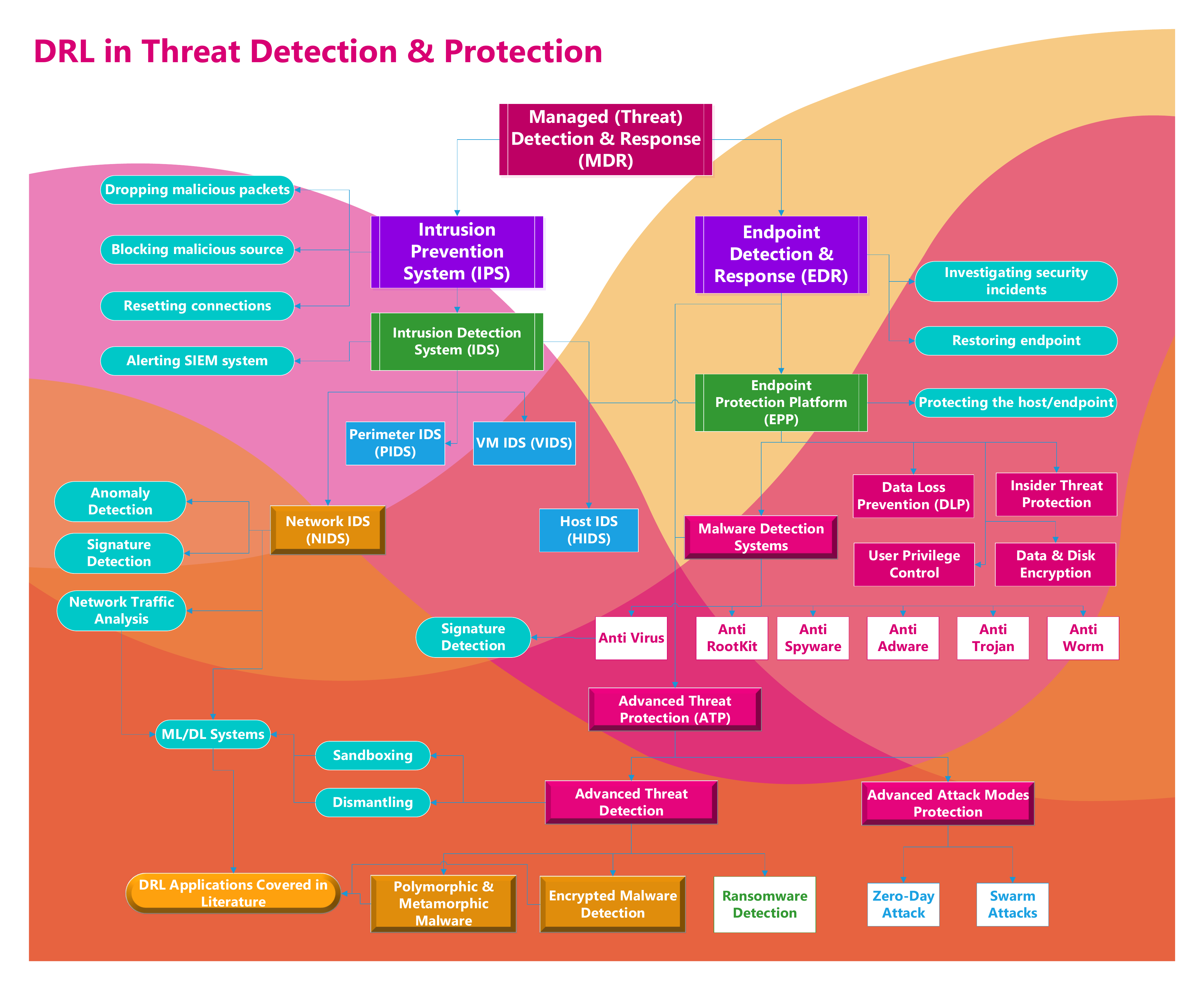}
    \caption{Taxonomy of Threat Detection \& Protection Systems in Cybersecurity and scope for DRL}
    \label{fig:mdr-taxonomy}
\end{figure}

Erstwhile, anti-virus and network firewall systems were deemed sufficient for protection against network and host-based threats. But with time the nature and potency of threats have changed, and so have the defences against them. Now the threats do not arise and attack in isolation, and therefore defending against them requires an integrated and managed (threat) detection and response system (MDR). As shown in figure \ref{fig:mdr-taxonomy}, MDR combines both the sub-components for network security and prevention (IDS/IPS/IDPs) and the endpoint-based security systems (EDR/EPP). MDR can also triage the threats detection and discovery across these different subsystems to provide capabilities to perform threat forensics. Next, we describe the network security and endpoint security-based sub-systems and the scope of DRL in these.

\subsection{Taxonomy of Intrusion Detection \& Prevention Systems}\label{sec:taxonomy-ids}
 On the network protection end, the need for dedicated network log and anomaly detection components were felt. Vendors offered both software-based and hardware solutions to suit unique needs and budget requirements. Such integrated network threat detection system is generically now called a Network-based Intrusion detection System (NIDS) \cite{architecture-nids}. By definition, such NIDS protected an entire local network or sub-network. But similar network threat detection needs were also felt broadly at the level of the network perimeter, individual applications (contained in a VM or container), and at the individual host level. Therefore, taking clues from the NIDS and adding to it platform-specific detection resources and techniques, IDS for the perimeter (PIDS), VM (VIDS), and hosts (HIDS) were also developed \cite{ids-review}. 
 Of these, the more popular research coverage is the NIDS followed by the HIDS \cite{hids-survey}.
 The PIDS and the VIDS mostly use scaled/modified versions of NIDS and hence are not extensively covered in the research literature.
 HIDS resides on the host machine and interacts with the OS  to determine malicious activities or processes by analysing various host activities like system calls, apps, and file access logs, etc.
 The NIDS has 2 components; one based on signature matching which provided definite signals; and another based on analysis of log files and configurations, which provided statistical trends and anomalies which need to be interpreted into signals of distinct types. Where a threat detected from a definitive analysis like signature matching could be immediately blocked by the system, others discovered based on statistical methods were relayed as events to a Security Information and Event Management (SIEM) system, monitored by domain experts who can do additional analysis to confirm an attack or malicious network traffic and trigger appropriate actions. Popular signature matching algorithms used in a NIDS is based on the Aho-Corasick algorithm \cite{pipelined-aho-corasick}. The Intrusion Prevention System, as shown in figure \ref{fig:mdr-taxonomy} includes sub-systems to automate and actively block some such threats. Such systems are called the Intrusion Prevention Systems (IPS) when they exist as a separate layer, or an Intrusion Detection \& Prevention System (IDPS) when these exist as a composite system.

\subsection{Taxonomy of Endpoint Detection \& Protection Systems}\label{sec:taxonomy-endpoint}
On the host protection end, the anti-virus system was the predominant mechanism. But this failed to prevent many evolving threats. Some initial host protection systems combined both the Host Intrusion Detection System (HIDS) \cite{architecture-nids} and the anti-virus system. Like the NIDS, the HIDS \cite{hids-survey} also uses a monitoring-based approach to generate threat detection. But instead of monitoring the live network traffic and the resultant generated log files like the NIDS, the HIDS would monitor the running processes, changes in the operating system (OS) registry, and other configurations. Instead of sending the events to a SIEM system, the changes are blocked until admin privileges are granted. The anti-virus system as used in earlier generation host protection modules used to be predominantly based on signature matching techniques and hence required a periodical update to its signature repository to detect any new threats. Since malware are of diverse types ranging beyond just viruses, therefore soon such anti-virus systems started including signatures for other threats like the Trojans \cite{trojan-survey}, Worms \cite{worms-survey}, Spyware, Adware, etc. As shown in figure \ref{fig:mdr-taxonomy}, modern Endpoint Protection Platforms (EPP) \cite{epp-gartner-magic-quadrant} as used by many enterprises today, also includes one or more components listed below beside the malware detection system.
\begin{itemize}
    \item  Data Loss Prevention (DLP): This is used to automatically identify documents and records containing personal and confidential information and prevent their accidental leakage and enforce policies around its usage and archival.
    \item Insider Threat (IT) Protection: This component detects any malicious intent by an internal employee to steal or leak confidential data/credentials of critical systems of an enterprise.
    \item Data \& Disk Encryption: This module encrypts the data and the disk to mitigate impacts arising from asset thefts.
    \item User Privilege Control: This module assists in enforcing different policies based upon the access privilege of different users.
\end{itemize}
In wake of the advanced threats arising from advanced second-generation malware modern EPP platforms also comprise an Advanced Threat Protection (ATP) sub-system \cite{atp-microsoft-apress}, which may consist of advanced ML/DL-based malware detection and protection components.

\section{DRL for Network Intrusion Detection Systems (NIDS)}\label{sec:drl-nids}

NIDS monitors maliciousness in the network traffic in a host agnostic manner. These are included in the larger umbrella of managed (thread) detection and response systems to provide comprehensive protection coupled with endpoint protection systems. Since these systems are host-agnostic they cannot leverage many popular host-specific file detection techniques and rely on analysis of the network statistics and binary packet data and header information. Before managed NIDS systems, network audit data were manually screened and analyzed to detect any malicious activity or a possible attack. But such methods were not scalable with the increasing network sizes, which led to the popularity of managed NIDS. NIDS analyses multiple sources ranging from application traces, user command data, and network packets to detection signals of attack or malicious payloads. 

Of late, the two predominant methods that the NIDS used to detect threats were \textit{signature} based identification \cite{pipelined-aho-corasick}, called the \textit{SNIDS} and (\textit{s}tatistical) anomaly detection \cite{nids-anomaly-survey}, called the \textit{ANIDS}. SNIDS methods rely on a database of pre-extracted signatures from different network payloads and a corresponding label for each extracted signature. Such methods are very efficient but often could not detect similar threats in the absence of an exactly matching signature. Also, such methods are useful mainly for malicious payload detection. The ANIDS on the other hand extracts patterns of different statistical measures to detect any shift from normal network activity and flags such an event. Such events are relayed to a Security Information and Event Management (SIEM) system, monitored by domain experts who can do additional analysis to confirm an attack or malicious network traffic and trigger appropriate actions.

\begin{figure}[htb!]
    \centering
    \includegraphics[width=\linewidth]{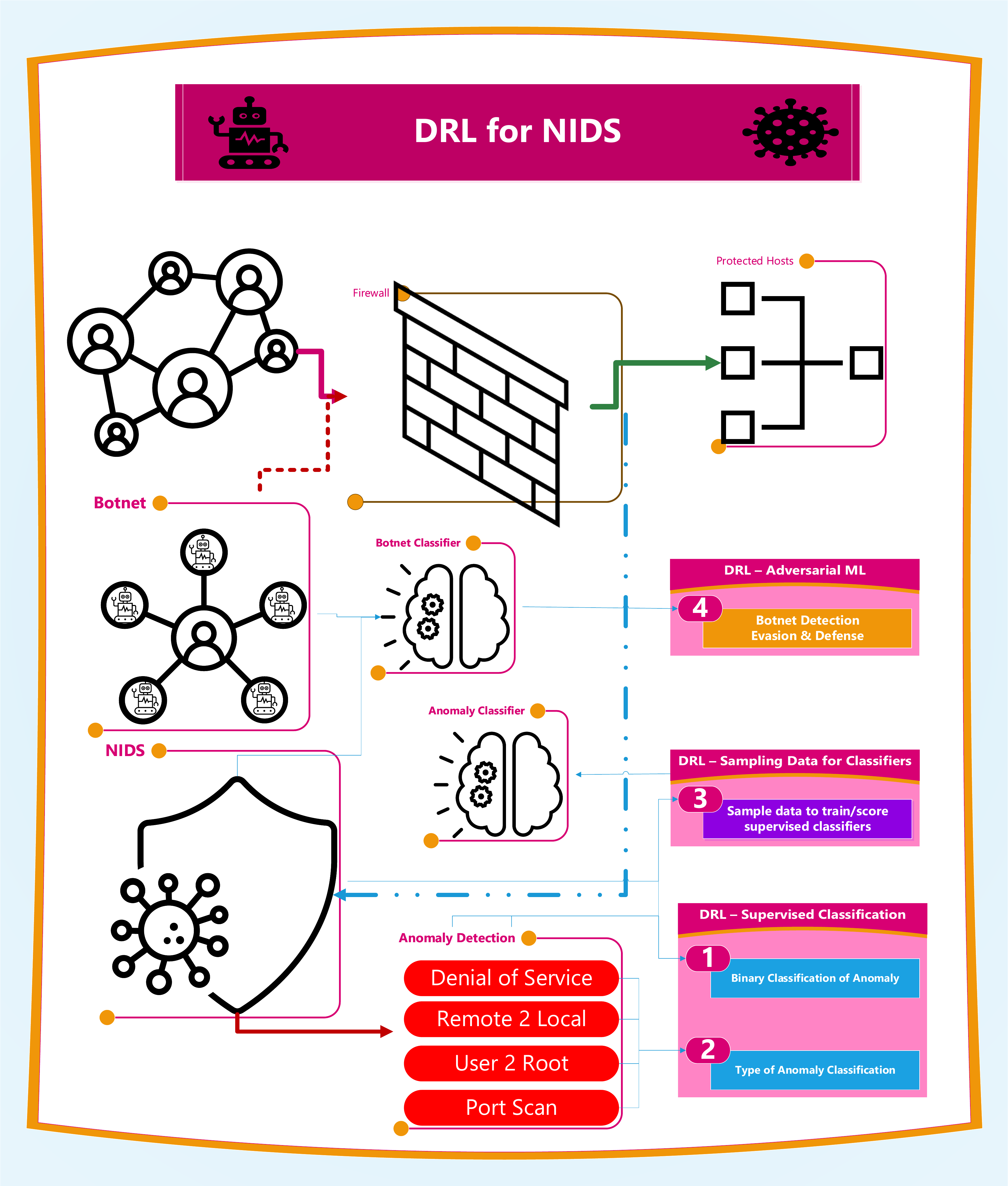}
    \caption{Application of DRL in NIDS}
    \label{fig:drl-nids-applications}
\end{figure}

The size of the network that a managed NIDS monitors is large and depending upon the strictness of threshold to any security event such ANIDS could generate high frequency and volume of detection/ identification events. When ANIDS is a part of an overarching IPS/IDPS, additional rule-based blocking/ analysis/ throttling actions could be triggered. But an automatic blockage/throttling of network traffic based on statistical anomalies is sub-optimal and may impact the network speed and latency for critical services. Therefore, such ANIDS are often coupled with ML/RL/DRL-based or hybrid systems. Such ML/RL-based systems could also be part of the active management done by IPS instead of the IDS. Recently DRL has also become popular for detecting an anomaly in the ANIDS systems. We review some recent art in this regard as shown in figure \ref{fig:drl-nids-applications} and in the following sub-sections, grouped by the objective of using DRL in ANIDS.

\subsection{DRL for anomalous network traffic (binary) classification}\label{sec:drl-anids-binary-classifier}
As shown in figure \ref{fig:rl-vs-drl}, the DRL consists of a DL approximator. In value-based DRL (figure \ref{fig:value-vs-policy-drl}), this estimator comprises a DL model architecture to estimate the state (\textit{V}) or state-action (\textit{Q}) value. Though DRL, as opposed to DL are conceived for tasks with sequential attribution for rewards, if we ignore this aspect, and map the instantaneous reward directly to a softmax function, then the DRL essentially works as a DL algorithm with an episode of unity size, and no accumulation or discounting of rewards for past attribution. 

Some arts that use DRL as a similar supervised DL learner are as follows:
\begin{itemize}
    \item Gülmez et. al. \cite{nids-drl-as-binary-classifier-gulmez} discussed a DRL-based approach for network intrusion detection and evaluated it on NSL-KDD \cite{nsl-kdd-dataset} and UNSW-NB15 \cite{unsw-nb15-dataset} datasets, which are the two most commonly used standardized datasets of network-anomalies. The results demonstrated that the DRL method achieved an F-scores of $\ge 0.96$ in both datasets. The DL approximator used here is a Multi-Layer Perceptron (MLP) based Deep Neural Network (DNN) \cite{sewak-OverviewDNN}. As with any DNN, the effectiveness of the DRL model here also was significantly affected by the structure of the DNN.
    \item Feng-Hsu et. al.\cite{nids-drl-as-binary-classifier-hsu} also used the same datasets (NSL-KDD \cite{nsl-kdd-dataset} and UNSW-NB15 \cite{unsw-nb15-dataset}) and also proposed a DRL-based classification approach for ANIDS. But they adopted an alternate testing approach, and in addition to using a standardized dataset for evaluation, they evaluated their system on a real campus NIDS as well, which had exponentially a larger network traffic flow. They also compared their DRL-based classification approach with three other ML-based classifiers (based on Random Forest, Support Vector Machine, and Multi-Layer Perceptron).
    \item Kamalakanta Sethi et. al. \cite{DRLNIDS3} presented a \textit{context-adaptive} IDS using DRL. Their main intention of replacing a DL-based classifier with a DRL-based classifier was to achieve a balance between detection accuracy and False Positive Rate (FPR). They noted that balancing between the accuracy and the FPR (robustness) using classical ML techniques is challenging, and by using DRL this could be achieved.
    
    Since all the 3 papers have worked with the same datasets, we can compare the results for each dataset. Here is a short comparative summary of results from these work on the two standardized datasets that all of these use.
    \begin{itemize}
        \item Evaluation of NSL-KDD dataset \cite{nsl-kdd-dataset}: Though the model proposed by Gülmez gave a 97\% accuracy with a precision of 98\% and a recall of 96\%, and the model used by Feng-Hsu gave an accuracy of 91.4\% their results are not directly comparable as they did not mention their FPR and for such classification accuracy alone could not be the basis of comparison. Feng-Hsu also claimed that DRL based approach gave better results when compared with the three classical ML techniques used and presented in their papers (RF, SVM, and MLP-DNN). The DQN based model used by Kamalakanta et.al., though had a reduced accuracy rate of 81.80\% on the dataset (when compared to the other two models), was able to reduce the FPR to 2.6\%.
        \item Evaluation of UNSW-NB15 dataset \cite{unsw-nb15-dataset}: The model proposed by Gülmez, again gave higher (but not directly comparable) accuracy of 96\%, at a 95\% precision, and a recall of 97\%. Like the results on NSL-KDD, the model used by Feng-Hsu gave a balanced result of 92\% across all the evaluation metrics on UNSW-NB15. The results presented by Kamalakanta on the other hand again had a reduced accuracy rate of 85.09\% and the False Positive Rate of 3.3\%. 
    \end{itemize}
\end{itemize}

\subsection{DRL for anomalous activity type classification in network traffic}\label{sec:drl-anids-multinomonal-classifier}

Most DL classifiers could classify both binary and multi-class problems. Therefore, while using DRL predominantly as a DL-based classifier, the detection type is only limited to the availability of classes in the standardized dataset. The popular NSL-KDD dataset \cite{nsl-kdd-dataset} for ANIDS, comes in various sizes, and difficulty levels for two types of classification problems; a binary class dataset to predict the presence of anomalous activity and a 5-class dataset, which has four types of anomalous activities, and the fifth class representing normal traffic. The arts in this sub-section cover identification of one or more of these types of anomalies. These are listed below:

\begin{itemize}
    \item Ekachai Suwannalai et. al. \cite{NIDSclassification} presented a DQN based approach for a five-label classification problem and compared the results with two other ML-based approaches, the Recurrent Neural Network (RNN) and the Adversarial Reinforcement Learning with SMOTE (AESMOTE). The model was trained on two standardized datasets, the NSL-KDD and the KDDTest+. They demonstrated that this approach gave superior performance than the other two non-DRL approaches.
    \item Yandong et. al. \cite{NIDSclassification2} used a DRL based framework which was based on an actor-critic based approach, called the Deep DPG (DDPG) (a policy-based method mention in section \ref{sec:about-drl}). The state space consisted of these 8 features namely the number of bytes and packets transmitted and those received, and the switch port number, etc. They used a conditional reward function as shown in equation \ref{eq:anomaly-binary-reward-function}:

    \begin{equation}
        \text{Reward} = 
        \begin{cases}
            -1 & \text{Load}_s > U_s\\
            \lambda p_b + (1-\lambda)(1-p_a) & \text{Load}_s \le U_s
        \end{cases}
        \label{eq:anomaly-binary-reward-function}
    \end{equation}
    
    They claimed that the proposed algorithm was capable of learning efficient mitigation policies and could mitigate an adverse Distributed Denial of Service (DDoS) attack in real-time. They also tested their framework on five other scenarios with different attack dynamics to prove the robustness of their system.
\end{itemize}

\subsection{DRL for Sampling Anomalies}\label{sec:drl-anids-sampling-data}

There is more that DRL can do as compared to DL, especially for scenarios that involve sequential decision-making beyond simple prediction on a static record. The art in this section covers approaches where RL/DRL are used to assist an ML/DL-based classification system instead of replacing it. This is done by selecting either the candidate input dataset for training or for scoring with these classifiers.
 
%  Another mechanism of implementing RL/ML based techniques is in layered/tandem detection. In one such tandem detection approach as suggested by Deokar and Hazarnis \cite{ids-log-files-rl}, the ML components are trained to detect anomalies in any given log file, whereas the RL component selects and prioritize the log files in which the anomalies need to be detected \cite{ids-log-files-rl}. 
 
Lopez-Martin et al. \cite{drl-ids-sampling} used DRL to selectively sample the training data used for training the Ml-based anomaly detection models. They used the four most popular DRL agent algorithms the DQN, the DDQN, the Policy Gradient (PG), and the Actor-Critic (AC). Of these, the DDQN gave the best results with all metrics including the accuracy, F1 score, precision, and recall $\approx 89\%$. They also benchmarked their DRL model's performance against ML-based techniques like Logistic Regression, SVM, Random Forest. The result indicated that where the DRL-based methods had improved accuracy, F1 score, and recall but algorithms like Adaboost generated better precision of $97\%$.

\subsection{DRL for botnet detection evasion}\label{sec:drl-anids-botnet-evasion}

A botnet \cite{understanding-botnet} is a malware-infected computer network that is under the control of a master attack perpetuating machine known as the bot-herder. With such a sophisticated network of malicious systems, the bot-herder is capable of carrying out a massive, coordinated attack. Common botnet attacks include DDoS, financial breaches, Email spams, etc. Therefore, the identification of such botnets in the network becomes very crucial but is also very challenging. 
As per work by Rajab et. al. \cite{understanding-botnet} up to 27\% of connections and 11\% of DNS domains are infected by a botnet. Contemporary botnets use evasion techniques to avoid detection by even potent ML/DL-based ANIDS. Therefore, recently DRL is being used to understand how botnets could evade detection from ANIDS and then bolster defences against such evasions. Next, we present some work that has used DRL in the area of botnet detection-evasion, such that effective defences could be created against a real evading botnet. 

Giovanni Apruzzese et. al. \cite{drl-adversarial-learning-botnet-evasion} suggested that a more resilient and robust botnet detection system can be developed through adversarial training. Two datasets were used to train the model, namely the CTU and BOTNET which trained two DRL agents based on DDQN and Deep-SARSA. The agent in their DRL system changes the binary sequence of a malicious network traffic flow with one of the available sequences of binaries until the embedded detector could not detect it as malicious.
The results showed that by contaminating the training set by 1\%, the evasion rate increased by 25\% approximately. 

Where Giovanni Apruzzese et. al. \cite{drl-adversarial-learning-botnet-evasion} worked on action-selection based policy, Di Wu et. al. \cite{drl-botnet-ml-evasion} proposed an action-elimination based policy for a DQN agent. They used a dataset made by combining benign samples from IOST 2010 dataset and botnet samples are taken from the \textit{Malware Capture Facility Project}. They gave an instantaneous reward of 10 units for a successful evasion. Interestingly, no penalty was given to the agent on failed evasion. Their action also included a simple pre-determined set of modifications to the binary traffic. Their results showed an evasion ranging from 70\% to 80\% on various botnet samples (namely the Nerris, Rbot, Zeus, and Geodo), which was significantly higher than the available baseline.

\section{DRL for Endpoint Detection \& Protection Systems} \label{sec:drl-endpoint}

An endpoint is the last line of defense for any host machine/device. Also, this is the only component that sees the file in its entirety as an executable program compatible with the host's operating system (OS), and hence offers opportunities for complex feature extraction and processing. Therefore, this is the most matured component from the perspective of the application of ML/DL/DRL techniques. 

With the evolving malware threats from advanced second-generation malware and ransomware, there is little that NIDS could do for their detection. Also, many such malware postpone any malicious activity that could be captured by a dynamic analysis system. Second-generation malware could use techniques like obfuscation \cite{ObfuscationTechniques} to change its structure and evade classical anti-virus programs. Therefore, the ATP-based systems, invariably use some form of ML for advanced threat detection and protection.

\begin{figure}[tb!]
    \centering
    \includegraphics[width=\textwidth]{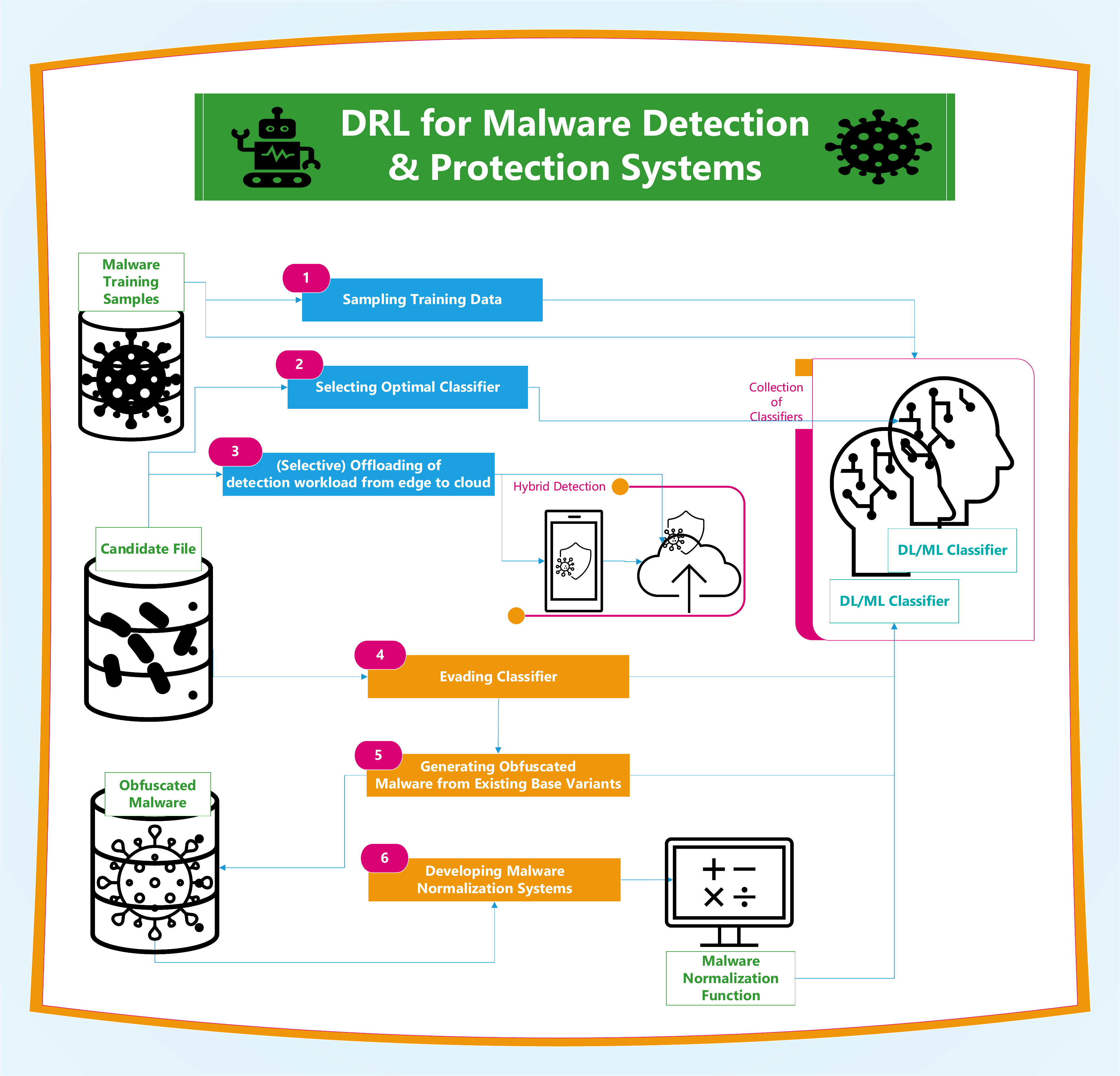}
    \caption{Applications of DRL in Malware Detection \& Protection Systems}
    \label{fig:drl-malware-detection-applications}
\end{figure}

Also, the use of DRL in endpoint security is much more advanced than for NIDS as covered in section \ref{sec:drl-nids}. There are two broad ways that DRL is used for ATP in an EDR/EPP as shown in figure \ref{fig:drl-malware-detection-applications}, these are enumerated below and detailed in the following subsections.

\subsection{DRL for assisting existing classification system} \label{sec:edr-assisting-classification}

In this subsection, various DRL applications are discussed to enhance the efficiency or effectiveness of classification by an existing ML/DL-based malware classification system. These applications fall under the following three types as discussed next. 

\subsubsection{Sampling data for training a malware classifier:} \label{sec:epp-sampling-data}

The SOTA ML and DL-based classifiers require a large amount of data on which they can train and learn. The supervised models are trained on a standardized network-anomaly dataset and learn to extract the best features from these datasets. The files, however, cannot be scanned indefinitely online by the systems and their execution needs to be halted by the heuristics employed by the engine somewhere.

A DRL-based supervised learner was proposed by Yu Wang et. al. \cite{DRL-sample-data-training} to determine when the emulation process should stop optimally based on the state information generated. Many Personal Executable (PE) files produced from Microsoft's anti-malware engine were used as the dataset to train the model. The agent can perform two types of actions continue (C) and halt (H) and was rewarded based on 2 criteria: firstly, based on the sequence lengths of the emulation, the shorter the sequence, the higher the reward (so that the agent tries to halt the process faster) and secondly the prediction accuracy.

The results from their work indicated that the DRL model was able to halt the execution of around 91\% of samples on which the model was tested. The true positive rate (61\% improvement) and false positive rate (1\% improvement) also improved when compared to the best baseline model presented by Ben Athiwaratkun et. al. \cite{baseline-classifier}

\subsubsection{Selecting Optimal Classifier for a candidate detection} \label{sec:epp-classifier-selection}
The modern-day threat detectors and classifiers use multiple detectors to increase their detection rates. Such models could be scaled horizontally using cloud infrastructure to reduce the computation time even with multiple candidate detection. The idea of using selected classifiers for candidate detection, and subsequently querying additional malware detectors was suggested, instead of using an ensemble of all the classifiers, and then aggregate their predictions. This approach may reduce the overall computation cost of a candidate file’s inference. But such targeted detection requires intelligent decision-making dynamically based on the incoming candidate’s file’s features. This is where DRL can help.

Yoni Birman et. al. \cite{costeffective} presented a DRL-based approach to implement a `cost-aware' malware detection approach to maintain a balance between first the need to have multiple ML classifiers for sophisticated detection and second for total computation cost for a candidate detection. They proposed to have such a cost-aware mechanism as a part of a Malware Detection as service (MDaaS) approach. Their system used an Actor-Critic agent and an action-space denoted by the array as in equation \ref{eq:action-yoni}, where \textit{N} is the number of detectors, \textit{M} is the Maximum number of detectors per action, and \textit{c} the binary class for the file.
\begin{equation}
    c+\sum_{i=1}^M \binom{N}{i}
    \label{eq:action-yoni}
\end{equation}

Further, as a reward, the correct classifications are assigned a fixed reward, and a penalty proportional to the computational cost involved during the classification was imposed. This helped in learning the behavior of not scoring against the classifiers that will have little or no contribution as a misclassification from these will incur a penalty for the DRL.

Their system was successfully able to maintain a similar level of accuracy as obtained by aggregating inferences from all available classifiers and reduced the computational cost by around $87\%$. They benchmark their system using a large number of the balanced dataset of PE files collected from the security department of a large organization and using three prominent cloud architectures namely the on-premise, AWS EC2, and AWS Lambda. 

\subsubsection{Selective offloading of edge detection workloads to cloud based detection} \label{sec:epp-cloud-offloading}
Mobile endpoints offer greater challenges as compared to PC and Servers. This is because of the availability of limited computation power for neural processing with any ML/DL-based system and battery/power constraints. Therefore, these systems invariably need to decide the optimal state of balance between edge detection and server/ cloud offloading of candidate detection. This is again an area where DRL can help.

In this regard, Xiaoyue Wan et. al. \cite{offloading} suggested a DQN based approach (and a hotbooting-Q strategy) to determine a strategy of selectively offloading detection from the endpoints on edge devices to a cloud-based detection setup. This is like the Q-Learning-based strategy proposed by Yanda Li et. al. \cite{Qlearningbasedoffloading}, however, Yanda et.al.'s method resulted in convergence, especially for larger networks. The use of DQN instead of classical Q-Learning resulted in faster convergence along with the higher accuracy, which was 24.5\% more than the traditional Q-learning-based approach as proposed by Yanda.

\subsection{DRL for adversarial attacks on existing classifiers and subsequent defense} \label{sec:edr-adversarila-ml}

In this subsection, we will discuss the applications of DRL to generate adversarial samples to evade an existing malware classifier. Metamorphic malware could use techniques like obfuscation \cite{ObfuscationTechniques} to similarly evade a classifier, though in a more organic manner than conventional adversarial-ml. We therefore also include arts to generate metamorphic malware and DRL-based techniques to defend against them also in this section. These applications fall under the following three types as discussed next.

\subsubsection{Evading malware classifier:} \label{sec:atp-evading-classifier}

 Advances in adversarial-ML have proved that a DL-based classifier is very sensitive towards samples with adversarial noise. Such DL-based classifiers could be easily evaded by strategically adding noise in the candidate file’s feature to lower the overall detection probability sufficiently to reverse the detection. These types of attacks are known as score-based attacks in adversarial-ml and are widely covered in DL literature. Earlier Generative Adversarial Networks (GAN) based techniques were used for this effect, but recently DRL based techniques are becoming popular.

% Weilin Xu et.al. \cite{PDFMalware} used two \textcolor{red}{PDF???} malware classifiers PDFrate and Hidost \textcolor{red}{these are not popular terms so explain}, and revealed that both of them were vulnerable to the attacks because they employed features which were non-robust and were easily manipulated without disrupting the desired malicious behaviour. Gross et. al. \textcolor{red}{cite???} used the DREBIN dataset (for Android Malware) for creating malware classifiers and then used a gradient based approach to construct highly-effective attacks to evade the classifiers. \textcolor{red}{why non-DRL approaches are coming this section? Are these the prelude to the DRL paper? if so there is no connection explicitly mentioned, these looks random summaries.}

Anderson et. al. \cite{evadingML} used a DQN based agent for malware evasion. The action space of their setup consisted of included basic adding, creation, and manipulation-based actions; and they used a binary unity reward for a successful evasion. They demonstrated a median evasion rate of $72\%$ using their method. 

In \cite{malware-drl-evading-classifier} Fang Z et. al. also used a DQN agent system to evade anti-malware engines and named it DQEAF. The DQN agent is trained on samples that are screened as malicious using the online VirusTotal service. Four such agents are trained on four different families of malware under the Win32 platform including `Backdoor’, `Trojan’, `Worm’, and `Email’.  The agent would append randomly generated bytes to the file features, libraries, and sections to the sample PE file to change their signatures. They also used a binary unity or 0 reward scheme as Anderson et.al., but with a different calculation for reward as shown in equation \ref{eq:evading-classifier-reward}
    
    \begin{equation}
        r_t = 20^{(-\text{TURN}-1) / \text{MAXTURNS}} \times 100
        \label{eq:evading-classifier-reward}
    \end{equation}
    
The RL learning episode would terminate on successful evasion. The results showed a slightly higher evasion success rate of $75\%$. The similarity in results between the work of Fang et.al. and Anderson et.al. is expected due to the similar algorithm used on a similar dataset but is not comparable as Fang et.al. did not use a standardized dataset.

\subsubsection{Generating obfuscated malware:} \label{sec:atp-generating-malware}

Advanced Threat Protection (ATP), is one of the most powerful features for any Endpoint detection/protection system. As shown in figure \ref{fig:mdr-taxonomy}, the ATP intends to protect against next/second generation threats like oligomorphic, polymorphic, and metamorphic malware. Of these, metamorphic malware is the most dangerous. Metamorphic malware could change their code body using techniques like obfuscation, and in doing so they could virtually infect every connected host in the entire network evading all types of current generation detection techniques including those based on sophisticated DL-based detection mechanisms. 
Some of the popular malware obfuscation techniques used include dead/junk code insertion, register reassignment, subroutine reordering, code transposition, etc. \cite{MOT} So far, the threat from these systems, though was high on impact or `value at stake' perspective, not so much from a `probability of occurrence'. This is because such malware is overly complex to develop.

Sewak et. al. \cite{sewak-ubicomp-doom}, showed how an adversarial-ML-based DRL system, using PPO agents could be used to generated obfuscated and metamorphic malware. Their system could be used to create such metamorphic malware of most of the existing non-metamorphic malware currently available in abundance. They created multiple PPO-based DRL agents as an adversary to ML-based malware detectors. There is a similar coverage of DRL as an adversarial to ML/DL-based classifiers in NIDS in section \ref{sec:drl-anids-botnet-evasion} for Botnet evasions. The techniques covered in this section are based upon popular value-based DRL agents like the DQN. As highlighted by Sewak et.al., the problem of obfuscation even at the opcode level; which is the highest abstracted level at which the functionality of a metamorphic malware could be preserved; is too complex to be effectively and efficiently solved by such popular DRL agent algorithms. This is because when the problem is converted into an equivalent single-task MDP, it leads to an extremely high cardinality action space, for which the existing popular DRL agent algorithms as we covered in section \ref{sec:drl-anids-botnet-evasion} do not scale well.

Interestingly, Sewak et.al. also argued that existing art in adversarial-ml especially the ones based on gradient techniques like GANs should not be used in the area of malware detection and protection. Some work covered in section \ref{sec:drl-nids} highlighted the motivation that some DL-based models become resilient to noise when trained on synthetically generated adversarial samples. Sewak et.al. explained that in endpoint protection, that the real threat is from malware that is compatible with the underlying host platform and could perform the desired malicious activities on that platform. Since synthetically generated perturbations from adversarial-ml-based systems like GAN (\cite{IDSGAN}) could not be compiled to host-compatible files, these mechanisms do not work for malware detection. Also, some research with GAN in security has pointed out that gradient-based adversarial samples do not work effectively in real life even if they could be made to evade another DL classifier \cite{MalGAN}. Sewak et.al. also argued that for malware generation only the techniques that preserve file functionality should be used as otherwise it is not guaranteed that the resulting file could behave as genuine malware. Because of these constraints there existed no work to generate a metamorphic malware, and this is the first work covered in literature to accomplish this. 

\subsubsection{Developing malware normalization systems} \label{sec:atp-malware-normalization}

In section \ref{sec:atp-generating-malware} we covered an obfuscated malware generation system as developed by Sewak et.al. In this, the authors highlighted that since their main objective is not to generate a metamorphic malware but defences against the same, also they indicated that re-training the existing malware classifier with the generated malware data that could evade the classifier will not be optimal. Therefore, further, they created a malware normalization system that could provide defense against metamorphic malware that uses similar obfuscation techniques to evade detection \cite{sewak-dsj-drldo}.

A malware \textit{normalization} system is a system that takes in an obfuscated executable file as input and uses de-obfuscation techniques to generate a common base form of the malware that is used to improve the detection rate of an existing malware detection system. All the different obfuscations of a given malware should be reverted to the same base variant for an existing malware that is trained on the malware's base variants. In this approach there is no need to re-train the existing malware classifier, thereby the malware's effectiveness against existing threats is not compromised by training it with adversarial sample data as pointed in \cite{GANissues}.

Sewak et.al. again used PPO to create this malware normalization system as well and evaluated it against obfuscated malware created from the Malicia dataset \cite{Malicia}. The results showed that \textbf{60\%} of the previously un-detectable obfuscated metamorphic malware variants were now detected by the same malware classifier after normalizing with their system.

\section{Conclusion}\label{sec:conclusion}
In this paper, we have presented the diverse and innovative applications of DRL in the field of cybersecurity threat detection and protection. With the increasing demand for more robust and capable defense systems, DRL has become a valuable tool in designing and enhancing the capabilities of modern NIDS and ATP systems. DRL is finding many unique applications in both intrusion and malware defense scenarios. It is enabling applications ranging from the ones that critically assist the development or selection of machine and deep learning algorithms to directly bolstering defences against advanced adversarial-ml attacks. DRL is also used in network anomaly detection as a replacement for a supervised learner, but the more empowering usage of DRL has been to form strategies for defense against modern and advanced adversarial attacks. DRL has also shown great promise in designing capable defences against advanced metamorphic malware. DRL itself is a trending research topic in AI and is continually being enhanced with more powerful and efficient algorithms and techniques. Therefore, we believe that it will empower many more unique and innovative applications of DRL in cybersecurity threat detection and protection, and therefore we believe that this is an interesting area for both cybersecurity and AI researchers to follow.

\bibliographystyle{unsrt}
\bibliography{main, ids, malware, ai, local}

\begin{thebibliography}{10}

\bibitem{drl-security-nguyen}
Thanh~Thi Nguyen and Vijay~Janapa Reddi.
\newblock Deep reinforcement learning for cyber security.
\newblock {\em arXiv preprint arXiv:1906.05799}, 2019.

\bibitem{statistica-epp-edr}
Shanhong Liu.
\newblock {Endpoint detection and response (EDR) and endpoint protection
  platform (EPP) market size worldwide from 2015 to 2020}.
\newblock
  \url{https://www.statista.com/statistics/799060/worldwideedr-epp-market-size/},
  2020.
\newblock [Online; accessed 27-June-2021].

\bibitem{drl-1-intro-rl}
Mohit Sewak.
\newblock Introduction to reinforcement learning.
\newblock In {\em Deep Reinforcement Learning}, pages 1--18. Springer, 2019.

\bibitem{drl-an-overview-li}
Yuxi Li.
\newblock Deep reinforcement learning: An overview.
\newblock {\em arXiv preprint arXiv:1701.07274}, 2017.

\bibitem{drl-survey-arulkumaran}
Kai Arulkumaran, Marc~Peter Deisenroth, Miles Brundage, and Anil~Anthony
  Bharath.
\newblock Deep reinforcement learning: A brief survey.
\newblock {\em IEEE Signal Processing Magazine}, 34(6):26--38, 2017.

\bibitem{Sewak-DRL}
Mohit Sewak.
\newblock {\em Deep Reinforcement Learning - Frontiers of Artificial
  Intelligence}.
\newblock Springer, 2019.

\bibitem{sewak-ValueDRL}
Mohit Sewak, Sanjay Sahay, and Hemant Rathore.
\newblock Value-approximation based deep reinforcement learning techniques: An
  overview.
\newblock In {\em 2020 IEEE 5th International Conference on Computing
  Communication and Automation (ICCCA)}, pages 379--384, 2020.

\bibitem{drl-10-policy-based-approaches}
Mohit Sewak.
\newblock Policy-based reinforcement learning approaches.
\newblock In {\em Deep Reinforcement Learning}, pages 127--140. Springer, 2019.

\bibitem{drl-8-dqn-ddqn}
Mohit Sewak.
\newblock Deep q network (dqn), double dqn, and dueling dqn.
\newblock In {\em Deep Reinforcement Learning}, pages 95--108. Springer, 2019.

\bibitem{DQN_Nature}
Volodymyr Mnih, Koray Kavukcuoglu, David Silver, Andrei~A. Rusu, Joel Veness,
  Marc~G. Bellemare, Alex Graves, Martin Riedmiller, Andreas~K. Fidjeland,
  Georg Ostrovski, Stig Petersen, Charles Beattie, Amir Sadik, Ioannis
  Antonoglou, Helen King, Dharshan Kumaran, Daan Wierstra, Shane Legg, and
  Demis Hassabis.
\newblock Human-level control through deep reinforcement learning.
\newblock {\em Nature}, 518(7540):529--533, February 2015.

\bibitem{Double_DQN}
Hado Van~Hasselt, Arthur Guez, and David Silver.
\newblock Deep reinforcement learning with double q-learning.
\newblock {\em CoRR}, abs/1509.06461, 2015.

\bibitem{Dueling_DQN}
Ziyu Wang, Tom Schaul, Matteo Hessel, Hado Van~Hasselt, Marc Lanctot, and Nando
  De~Freitas.
\newblock Dueling network architectures for deep reinforcement learning.
\newblock In {\em Proceedings of the 33rd International Conference on
  International Conference on Machine Learning - Volume 48}, ICML'16, pages
  1995--2003. JMLR.org, 2016.

\bibitem{AE-DQN}
Tom Zahavy, Matan Haroush, Nadav Merlis, Daniel~J Mankowitz, and Shie Mannor.
\newblock Learn what not to learn: Action elimination with deep reinforcement
  learning.
\newblock {\em Advances in Neural Information Processing Systems},
  31:3562--3573, 2018.

\bibitem{drl-13-deep-deterministic-policy-gradient}
Mohit Sewak.
\newblock Deterministic policy gradient and the ddpg.
\newblock In {\em Deep Reinforcement Learning}, pages 173--184. Springer, 2019.

\bibitem{TRPO}
John Schulman, Sergey Levine, Pieter Abbeel, Michael Jordan, and Philipp
  Moritz.
\newblock Trust region policy optimization.
\newblock In {\em International conference on machine learning}, pages
  1889--1897. PMLR, 2015.

\bibitem{PPO}
John Schulman, Filip Wolski, Prafulla Dhariwal, Alec Radford, and Oleg Klimov.
\newblock Proximal policy optimization algorithms.
\newblock {\em CoRR}, abs/1707.06347, 2017.

\bibitem{GAE}
John Schulman, Philipp Moritz, Sergey Levine, Michael Jordan, and Pieter
  Abbeel.
\newblock High-dimensional continuous control using generalized advantage
  estimation.
\newblock {\em arXiv preprint arXiv:1506.02438}, 2015.

\bibitem{architecture-nids}
R~Heady, G~Luger, A~Maccabe, and M~Servilla.
\newblock The architecture of a network level intrusion detection system.
\newblock {\em Office of Scientific and Technical Information, U.S. Department
  of Energy}, 8 1990.

\bibitem{ids-review}
Hung-Jen Liao, Chun-Hung {Richard Lin}, Ying-Chih Lin, and Kuang-Yuan Tung.
\newblock Intrusion detection system: A comprehensive review.
\newblock {\em Journal of Network and Computer Applications}, 36(1):16--24,
  2013.

\bibitem{hids-survey}
Robert~A. Bridges, Tarrah~R. Glass-Vanderlan, Michael~D. Iannacone, Maria~S.
  Vincent, and Qian~(Guenevere) Chen.
\newblock A survey of intrusion detection systems leveraging host data.
\newblock {\em ACM Comput. Surv.}, 52(6), November 2019.

\bibitem{pipelined-aho-corasick}
Derek Pao, Wei Lin, and Bin Liu.
\newblock A memory-efficient pipelined implementation of the aho-corasick
  string-matching algorithm.
\newblock {\em ACM Trans. Archit. Code Optim.}, 7(2), October 2010.

\bibitem{trojan-survey}
Mohammad Tehranipoor and Farinaz Koushanfar.
\newblock A survey of hardware trojan taxonomy and detection.
\newblock {\em IEEE Design Test of Computers}, 27(1):10--25, 2010.

\bibitem{worms-survey}
Darrell~M. Kienzle and Matthew~C. Elder.
\newblock Recent worms: A survey and trends.
\newblock In {\em Proceedings of the 2003 ACM Workshop on Rapid Malcode}, WORM
  '03, page 1–10, New York, NY, USA, 2003. Association for Computing
  Machinery.

\bibitem{epp-gartner-magic-quadrant}
Peter Firstbrook, Arabella Hallawell, John Girard, and Neil MacDonald.
\newblock Magic quadrant for endpoint protection platforms.
\newblock {\em Gartner RAS Core Research Note G}, 208912, 2009.

\bibitem{atp-microsoft-apress}
Vasantha Lakshmi.
\newblock {\em Beginning Security with Microsoft Technologies}.
\newblock Springer, 2019.

\bibitem{nids-anomaly-survey}
Monowar~H. Bhuyan, D.~K. Bhattacharyya, and J.~K. Kalita.
\newblock Network anomaly detection: Methods, systems and tools.
\newblock {\em IEEE Communications Surveys Tutorials}, 16(1):303--336, 2014.

\bibitem{nids-drl-as-binary-classifier-gulmez}
Halim~Görkem Gülmez and Pelin Angın.
\newblock A study on the efficacy of deep reinforcement learning for intrusion
  detection.
\newblock {\em Sakarya University Journal of Computer and Information
  Sciences}, 4:11 -- 25, 2021.

\bibitem{nsl-kdd-dataset}
Ghulam Mohi-ud din.
\newblock Nsl-kdd dataset.
\newblock \url{https://www.unb.ca/cic/datasets/nsl.html}, 2017.
\newblock [Online; accessed 27-June-2021].

\bibitem{unsw-nb15-dataset}
Wells David.
\newblock Nsl-kdd datasets.
\newblock \url{https://www.kaggle.com/mrwellsdavid/unsw-nb15}, 2019.
\newblock [Online; accessed 27-June-2021].

\bibitem{sewak-OverviewDNN}
Mohit Sewak, Sanjay~K Sahay, and Hemant Rathore.
\newblock An overview of deep learning architecture of deep neural networks and
  autoencoders.
\newblock {\em Journal of Computational and Theoretical Nanoscience},
  17(1):182--188, 2020.

\bibitem{nids-drl-as-binary-classifier-hsu}
Ying-Feng Hsu and Morito Matsuoka.
\newblock A deep reinforcement learning approach for anomaly network intrusion
  detection system.
\newblock In {\em 2020 IEEE 9th International Conference on Cloud Networking
  (CloudNet)}, pages 1--6, 2020.

\bibitem{DRLNIDS3}
Kamalakanta Sethi, Sai Edupuganti, Rahul Kumar, Padmalochan Bera, and
  Y.~Madhav.
\newblock A context-aware robust intrusion detection system: a reinforcement
  learning-based approach.
\newblock {\em International Journal of Information Security}, 19, 12 2020.

\bibitem{NIDSclassification}
Ekachai Suwannalai and Chantri Polprasert.
\newblock Network intrusion detection systems using adversarial reinforcement
  learning with deep q-network.
\newblock In {\em 2020 18th International Conference on ICT and Knowledge
  Engineering (ICT KE)}, pages 1--7, 2020.

\bibitem{NIDSclassification2}
Yandong Liu, Mianxiong Dong, Kaoru Ota, Jianhua Li, and Jun Wu.
\newblock Deep reinforcement learning based smart mitigation of ddos flooding
  in software-defined networks.
\newblock In {\em 2018 IEEE 23rd International Workshop on Computer Aided
  Modeling and Design of Communication Links and Networks (CAMAD)}, pages 1--6,
  2018.

\bibitem{drl-ids-sampling}
Manuel Lopez-Martin, Belen Carro, and Antonio Sanchez-Esguevillas.
\newblock Application of deep reinforcement learning to intrusion detection for
  supervised problems.
\newblock {\em Expert Systems with Applications}, 141:112963, 2020.

\bibitem{understanding-botnet}
Moheeb Abu~Rajab, Jay Zarfoss, Fabian Monrose, and Andreas Terzis.
\newblock A multifaceted approach to understanding the botnet phenomenon.
\newblock In {\em Proceedings of the 6th ACM SIGCOMM Conference on Internet
  Measurement}, IMC '06, page 41–52, New York, NY, USA, 2006. Association for
  Computing Machinery.

\bibitem{drl-adversarial-learning-botnet-evasion}
Giovanni Apruzzese, Mauro Andreolini, Mirco Marchetti, Andrea Venturi, and
  Michele Colajanni.
\newblock Deep reinforcement adversarial learning against botnet evasion
  attacks.
\newblock {\em IEEE Transactions on Network and Service Management},
  17(4):1975--1987, 2020.

\bibitem{drl-botnet-ml-evasion}
Di~Wu, Binxing Fang, Junnan Wang, Qixu Liu, and Xiang Cui.
\newblock Evading machine learning botnet detection models via deep
  reinforcement learning.
\newblock In {\em ICC 2019 - 2019 IEEE International Conference on
  Communications (ICC)}, pages 1--6, 2019.

\bibitem{ObfuscationTechniques}
Chandan~Kumar Behera and D~Lalitha Bhaskari.
\newblock Different obfuscation techniques for code protection.
\newblock {\em Procedia Computer Science}, 70:757--763, 2015.

\bibitem{DRL-sample-data-training}
Yu~Wang, Jack~W. Stokes, and Mady Marinescu.
\newblock Neural malware control with deep reinforcement learning.
\newblock In {\em MILCOM 2019 - 2019 IEEE Military Communications Conference
  (MILCOM)}, pages 1--8, 2019.

\bibitem{baseline-classifier}
Ben Athiwaratkun and Jack~W. Stokes.
\newblock Malware classification with lstm and gru language models and a
  character-level cnn.
\newblock In {\em 2017 IEEE International Conference on Acoustics, Speech and
  Signal Processing (ICASSP)}, pages 2482--2486, 2017.

\bibitem{costeffective}
Yoni Birman, Shaked Hindi, Gilad Katz, and Asaf Shabtai.
\newblock Cost-effective malware detection as a service over serverless cloud
  using deep reinforcement learning.
\newblock In {\em 2020 20th IEEE/ACM International Symposium on Cluster, Cloud
  and Internet Computing (CCGRID)}, pages 420--429, 2020.

\bibitem{offloading}
Xiaoyue Wan, Geyi Sheng, Yanda Li, Liang Xiao, and Xiaojiang Du.
\newblock Reinforcement learning based mobile offloading for cloud-based
  malware detection.
\newblock In {\em GLOBECOM 2017 - 2017 IEEE Global Communications Conference},
  pages 1--6, 2017.

\bibitem{Qlearningbasedoffloading}
Yanda Li, Jinliang Liu, Qiangda Li, and Liang Xiao.
\newblock Mobile cloud offloading for malware detections with learning.
\newblock In {\em 2015 IEEE Conference on Computer Communications Workshops
  (INFOCOM WKSHPS)}, pages 197--201, 2015.

\bibitem{evadingML}
Hyrum~S Anderson, Anant Kharkar, Bobby Filar, and Phil Roth.
\newblock Evading machine learning malware detection.
\newblock 2017.

\bibitem{malware-drl-evading-classifier}
Zhiyang Fang, Junfeng Wang, Boya Li, Siqi Wu, Yingjie Zhou, and Haiying Huang.
\newblock Evading anti-malware engines with deep reinforcement learning.
\newblock {\em IEEE Access}, 7:48867--48879, 2019.

\bibitem{MOT}
Ilsun You and Kangbin Yim.
\newblock Malware obfuscation techniques: A brief survey.
\newblock In {\em 2010 International Conference on Broadband, Wireless
  Computing, Communication and Applications}, pages 297--300, 2010.

\bibitem{sewak-ubicomp-doom}
Mohit Sewak, Sanjay~K. Sahay, and Hemant Rathore.
\newblock {DOOM:} a novel adversarial-drl-based op-code level metamorphic
  malware obfuscator for the enhancement of {IDS}.
\newblock In {\em UbiComp/ISWC '20: 2020 {ACM} International Joint Conference
  on Pervasive and Ubiquitous Computing and 2020 {ACM} International Symposium
  on Wearable Computers, Virtual Event, Mexico, September 12-17, 2020}, pages
  131--134. {ACM}, 2020.

\bibitem{IDSGAN}
Zilong Lin, Yong Shi, and Zhi Xue.
\newblock {IDSGAN:} generative adversarial networks for attack generation
  against intrusion detection.
\newblock {\em CoRR}, abs/1809.02077, 2018.

\bibitem{MalGAN}
Weiwei Hu and Ying Tan.
\newblock Generating adversarial malware examples for black-box attacks based
  on {GAN}.
\newblock {\em CoRR}, abs/1702.05983, 2017.

\bibitem{sewak-dsj-drldo}
Mohit Sewak, Sanjay Sahay, and Hemant Rathore.
\newblock Drldo a novel drl based de obfuscation system for defence against
  metamorphic malware.
\newblock {\em Defence Science Journal}, 71(1):55--65, Feb. 2021.

\bibitem{GANissues}
Weiwei Hu and Ying Tan.
\newblock Generating adversarial malware examples for black-box attacks based
  on gan.
\newblock {\em ArXiv}, abs/1702.05983, 2017.

\bibitem{Malicia}
Antonio Nappa, M.~Zubair Rafique, and Juan Caballero.
\newblock The malicia dataset: Identification and analysis of drive-by download
  operations.
\newblock {\em Int. J. Inf. Secur.}, 14(1):15–33, February 2015.

\end{thebibliography}
\end{document}